\documentstyle[12pt]{article}
\begin{document}
\begin{titlepage}
\title{Is the gravitational action additive?}
\author{Dieter Brill\thanks{e-mail: brill@umdhep.umd.edu}\\
Department of Physics,
University of Maryland,\\
College Park, MD 20742\\
and\\
Geoff Hayward\thanks{e-mail:
hayward@phys.ualberta.ca}\\
Institute of Theoretical Physics,\\
412 Avadh Bhatia Physics Laboratory,
University of Alberta,\\
Edmonton, AB,
Canada T6G 2J1}
\date{PACS: 04.20.Cv, 04.60-m}
\maketitle
\begin{abstract}
The gravitational action is not always additive in the usual sense.
We provide a general prescription for the change in action
that results when different portions of the boundary
of a spacetime are topologically identified.
We discuss possible implications for the superposition law
of quantum gravity.
We present
a definition of `generalized additivity' which does
hold for arbitrary spacetime composition.
\end{abstract}
\end{titlepage}
\baselineskip=10.mm
\section{\label{S1}Introduction}
It is widely believed that, when a spacetime is cut up into several parts,
the actions for the parts,
considered separately,
sum up to yield
the action for the spacetime taken as a whole.
In fact, Hawking has argued~\cite{Hawking/centenary} that
additivity of the action in this sense is actually
{\em necessary}
in order to obtain the appropriate
superposition
behavior for quantum gravitational propagation
amplitudes.
Viewed from a broader perspective,
this form of additivity would
require that the action for a spacetime
(which need not be connected)
be invariant under topological identifications of isometric
surfaces on its boundary.

Yet, as has been
pointed out by Hartle and Sorkin
in the context of Euclidean Regge calculus~\cite{Hartle/Sorkin},
the gravitational action
is not always additive in this sense.
Here we consider the context of continuum general relativity
and provide a general prescription
for the `residue'; that is,
a prescription for the change in action that results when one
topologically
identifies
isometric surfaces on the boundary of a
spacetime.  We then argue that in order
to maintain `additivity', we must broaden
this concept to allow for a finite action associated
with certain identification
surfaces themselves.  In this broader sense, the action
is always additive.

In this paper we establish that
there is a sense in which the action is additive and
a sense in which it is not.
It is therefore useful to
adopt appropriate terminology at the outset.
If the action is
additive in the usual sense (e.g. the sense used both
by Hawking and by Hartle and Sorkin) we will
call it `invariant under identification of
boundary surfaces' or, equivalently,
`invariant under
spacetime
composition'.  In Section \ref{S4}, we will define
a broader sense of additivity, which
we refer to as `generalized additivity'.

Let us now consider the gravitational action for a spacetime
$\{g^{ab}, {\cal M}\}$ with a
boundary $\{h^{ab},B\}$,
of fixed intrinsic geometry.  When the
unit normal varies continuously over the
boundary, the
action is given by~\cite{York1972,Gibbons/Hawking},
\begin{equation}I={1\over 16\pi}\int_{\cal M}
(R-2\Lambda)\,(-g)^{1/2}d^4x
-\varepsilon{1\over 8\pi}
\int_B K(\varepsilon h)^{1/2}\,d^3x
+ C,\label{action}\end{equation}
where $\varepsilon=+1~(=-1)$
when the boundary is spacelike (timelike),
where $K_{ab}$ is the extrinsic
curvature of the boundary,
and where $C$ is some undetermined functional
of the boundary's intrinsic geometry.

When the boundary includes sharp 2--dimensional
`edges' or `joints' (i.e. 2--surfaces at which
the unit normal changes direction discontinuously)
there are additional corrections to
(\ref{action})~\cite{2,Brill,Hartle/Sorkin,o2}.
The effect of the joint's extrinsic geometry is simply
a finite additive contribution to the action.
This contribution
depends on whether the joint is
spacelike, timelike, or null (i.e. on whether the
square of its area form is positive, negative, or zero).
A timelike joint contributes to the action
\begin{equation}
{1\over 8\pi}\int \Theta\,(-\sigma)^{1/2}d^2x
+ {\cal C}[\sigma_{ab}]
\label{jointcont}
\end{equation}
where $\Theta$ is the angle between
the outward pointing boundary normals
on either side of the 2--surface\footnote{Here $\Theta$ is
considered positive if these normals are diverging, and negative
if they are converging. A similar convention determines the sign
of $\eta$ of equation (3) below.},
$\sigma_{ab}$ is the intrinsic two-metric
of the 2--surface, and ${\cal C}$ is an
arbitrary functional of $\sigma_{ab}$.
Spacelike joints contribute to the action,
\begin{equation}
{1\over8\pi}\int \eta\,(\sigma)^{1/2}d^2x
+ {\cal C}[\sigma_{ab}]
\label{jointcont2}\end{equation}
where $\eta$ is the local boost parameter.
Null joints make no contribution
to the action
(except possibly through ${\cal C}$),
essentially because their 2--volume vanishes.

Now consider the following general procedure
by which a spacetime can be topologically reconfigured.
Let $\{g^{ab},{\cal M}\}$ be a spacetime
(not necessarily connected) with boundary $\{h^{ab},B\}$
(Figure 1).
Construct from this spacetime a new
spacetime, $\{{g}^{ab},
\widetilde{{\cal M}}\}$ with boundary
$\{h^{ab},\widetilde{B}\}$,
by identifying topologically similar surfaces
$S_1$ and $S_2$ on $B$
to form a single surface $S$.
Assume that the surfaces to be identified
are isometric, but allow for a jump
in extrinsic curvature between them.
Let $\widehat{S}$ be the closure of $S$
and let $S_{\rm int}$ be the interior of $S$. Define
$\bar{S}\equiv S-S_{\rm int}$
to be the portion of the boundary of $S$ that
is included in $S$.  Similarly, define
$\bar{S}^\ast\equiv\widehat{S}-S$
to be the portion of the boundary of $S$ that
is not included in $S$.

The behavior of the action (\ref{action})
under spacetime composition depends critically on the 2--surfaces
$\bar{S}$ and $\bar{S}^\ast$. (If these vanish,
then the action is invariant under the composition.)
In general, these 2--surfaces may include portions
that are timelike, spacelike, or null.
Since null joints do not contribute to the action,
only the timelike and spacelike portions of
$\bar{S}$ and $\bar{S}^\ast$
could conceivably contribute to a change in the action
when $S_1$ and $S_2$ are topologically identified.

\section{\label{S2}Identifying
surfaces with timelike boundary}

For simplicity,
first focus on the case in which
the boundary of the identification surface $S$ is
entirely timelike,
in which $S=\widehat{S}$
(so $\bar{S}^\ast=0$), and in which the
`outer sides' of $S_1$ and $S_2$
are identified (see Figure 2(a)).

Now compare the total action before ($I$) and
after ($\widetilde{I})$
we identify
${S}_1$ and $S_2$.
After identification, a jump discontinuity in the
extrinsic curvature across $S_{\rm int}$ contributes
\begin{equation}
{\varepsilon\over8\pi}
\int_{S_{\rm int}}\left.K\right|_-^+(\varepsilon
h)^{1/2}\,d^3x\label{jump}\end{equation}
to the action, which is the same as the contribution due to $S_1$
and $S_2$ before identification
(because extrinsic curvature contributions
along $S_{\rm int}$ for $S_1$ and $S_2$
are defined with respect to oppositely oriented
unit normals). Thus we find
\begin{equation}
I-\widetilde{I}={1\over 8\pi}
\int_{\bar{S}}\left(
\Theta_1+\Theta_2-\Theta
\right)(-\sigma)^{1/2}\,d^2x
+C-\widetilde{C},\label{diff}
\end{equation}
where $\Theta_1$ and $\Theta_2$
are the angles between outward pointing normals
at $\bar{S}_1$ and $\bar{S}_2$ before identification.
Noting that
$\Theta=\Theta_1 + \Theta_2-\pi$
(see Figure 2(b), and Ref.~\cite{Hartle/Sorkin}),
we obtain,
\begin{equation}
I-\widetilde{I}=A[\bar{S}]/8+
C-\widetilde{C},\label{diff2}\end{equation}
where $A[\bar{S}]$ is the area of $\bar{S}$.
Clearly, the right hand side of (\ref{diff2})
does not vanish for arbitrary
$C$.

Define the `residue'
${\cal R}[S]$, associated with identifying surfaces
$S_1$ and $S_2$ to form a surface $S$,
as the
difference between the value of
the action
before identification and its value
after identification for
the case that all $C$'s vanish.
It is not difficult to extrapolate from the above
example to obtain a general prescription for the
residue associated with $S$ when $\bar{S}$ and $\bar{S}^\ast$
are everywhere timelike.
We obtain
\begin{equation} {\cal R}[S]=\lambda A[\bar{
S}]/8-\lambda
A[\bar{S}^\ast]/8\label{resid}\end{equation}
where $\lambda=+1$ if the outer
sides of $S_i$ have been identified
and $\lambda=-1$ if the inner
sides of $S_i$ have been identified.

A special class of the topological
identifications considered above occurs when
different portions of $B$
are identified at isolated
2--surfaces.  Such identifications can be of interest
in the treatment of horizon
thermodynamics and cosmic string dynamics.  In these
cases
the inclusion or exclusion
of an isolated 2--surface
can have important implications for the global
properties of the spacetime.
We can derive the residue associated with
this class of identifications
with the aid of equation (\ref{resid}) and
a simple limiting procedure.

For instance, reconsider the example discussed
above (i.e. with $S\equiv \widehat{S}$ and outer
sides of $S_1$ and $S_2$ identified)
in the limit that
$S$ collapses on some timelike 2--surface,
$J$ (see
Figure 3).
Take the topology of $S$ to be
$J\times \bar{{\sf I}}$ where $\bar{{\sf I}}$ corresponds to
a closed interval. Now let a coordinate
$\varepsilon$ parametrize the interval
so that $S$ can be decomposed into a
family of 2--surfaces, $J_{(\varepsilon)}$.
Suppose that $-\varepsilon_0<\varepsilon<+\varepsilon_0$
and take the limit $\varepsilon_0\to 0$.
We obtain,
\begin{eqnarray}
{\cal R}[J]&=&
A[J_{(-\varepsilon_0)}]/8+A[J_{(+\varepsilon_0)}]/8
\nonumber\\
&=&A[J]/4\label{residerat}.
\end{eqnarray}

To obtain the residues associated
with more complicated
topological identifications, one
can combine sequences of simple
topological identifications
at 3--surfaces and 2--surfaces of the form described above.
In such cases, a diagram algebra can be
a valuable aid (see Figure 4).

\section{\label{S3}
Identifying surfaces with spacelike boundary}

Let us now examine whether
spacelike portions of $\bar{S}$
contribute to the residue.
Again, suppose that a
spacetime $\{g^{ab},{\cal M}\}$ is topologically
reconfigured to form a new
spacetime $\{g^{ab},\widetilde{{\cal M}}\}$
by identifying boundary
surfaces $S_1$ and $S_2$ to form
a single surface $S$.
First, consider the special case in
which $S=\widehat{S}$ with
$\bar{S}$ entirely
spacelike.  Furthermore, for definiteness, suppose that
$S$ is spacelike and $B-S$ is timelike in the neighborhood
of $\bar{S}$ (see Figure 5).

In the 2-space orthogonal to $\bar{S}_1$,
let $n^a_1$ be the unit vector tangent to $S_1$ and pointing
outward;
let $u^a_1$ be the future pointing
unit vector tangent to $B - S_1$; and let
$\tilde{u}^a_1$ be the future pointing normal to $n^a_1$.
Similarly, let $u^a_2$ and $n^a_2$ be past and outward pointing
unit normals at $\bar{S}_2$.
Before identification we have~\cite{2},
$\eta_1=\sinh^{-1}{(u_1\cdot n_1)}$
and $\eta_2=\sinh^{-1}{(u_2\cdot n_2})$.
After identification we
have $\eta=\cosh^{-1}{(n_1\cdot n_2)}$.
Now express $u^a_2$ and $n^a_2$ in terms
of vectors on $\bar{S}_1$:
\begin{eqnarray}
u_2^a&=&-u_1^a\nonumber\\
n^a_2&=& n^a_1\cosh{\eta}+\tilde{u}^a_1\sinh{\eta}
.\label{vectors}\end{eqnarray}
Use (\ref{vectors}),
the definition of $\eta_2$,
and the fact that
$u_1\cdot
{\tilde{u}}_1=-\cosh{\eta_1}$,
to find
\begin{eqnarray}
\eta_2&=&\sinh^{-1}{}\left(u_2
\cdot n_2\right)\nonumber\\
&=&-\sinh^{-1}{}\left(u_1\cdot(
n^a_1\cosh{\eta}+\tilde{u}^a_1\sinh{\eta}
)\right)\nonumber\\
&=&-\sinh^{-1}{}\left(\sinh{\eta_1}\cosh{\eta}
-\cosh{\eta_1}\sinh{\eta}\right)
\nonumber\\
&=&\eta-\eta_1.\label{whew}\end{eqnarray}
Now an analysis parallel
to that done in the previous section yields
\begin{eqnarray}
I-\widetilde{I}&=&{1\over 8\pi}
\int_{\bar{S}}\left(
\eta_1+\eta_2-\eta
\right)(-\sigma)^{1/2}\,d^2x
+C-\widetilde{C}\cr
&=&C-\widetilde{C}
.\label{diff3}
\end{eqnarray}
Setting $C=0$, we have that {\em the residue associated
with $\bar{S}$ vanishes}.

Numerous other examples can be
constructed in which the identification
surface has a spacelike boundary.
For instance, one can
choose the identification surface
to be timelike rather
than spacelike.  Also, one can consider
cases in which the unit normals
at $\bar{S}_1$ and $\bar{S}_2$
are differently
oriented (e.g. all timelike or
all spacelike or some
other combination of spacelike and timelike).
By analysis parallel to that conducted
above, we find in each case that the residue
associated with the spacelike boundary
of the identification surface is zero.

\section{\label{S4}Composition invariance and the problem of superposition}
According to the path integral ansatz for quantum
gravity, the amplitude for propagation
from a state $|{\bf{g}}_i,S_i\rangle$
to a state $|{\bf{g}}_f,S_f\rangle$ is
\begin{equation}
\langle {\bf{g}}_f,S_f|{\bf{g}}_i,S_i\rangle=
\int {\cal D}[g]\exp{iI[g]}.\end{equation}
To obtain proper superposition behavior,
we require
\begin{equation}
\langle {\bf{g}}_f,S_f|{\bf{g}}_i,S_i\rangle=
\sum_{{\bf g}}\langle {\bf{g}}_f,S_f|{\bf{g}},S\rangle
\langle {\bf{g}},S|{\bf{g}}_i,S_i\rangle.
\label{super}\end{equation}
It is often claimed (see, for instance,
Ref.~\cite{Hawking/centenary})
that equation (\ref{super}) can
be satisfied only if the action is invariant under topological
identifications of boundary surfaces.
How then do the results of Sections \ref{S2}
and \ref{S3} affect the superposition behavior
of quantum gravitational propagators?

Consider first the most common case where the surfaces
$S_i$, $S$, and $S_f$ are spacelike, compact, and joined
at their boundaries by a timelike tube at large radius.
The regions to the future and past of $S$ then have a joint
at the boundary of $S$ that might result in a residue.
However, if $S$ is a spacelike hypersurface,
its boundary must also be spacelike.
 From the results of Section 3 we can now conclude that
in fact the boundary of $S$ will {\em not} contribute to the
residue. Thus, if we consider only propagation
between {\em spacelike} surfaces, the propagator will always
have the proper superposition behavior when $C=0$.
(Other choices of $C$ are also consistent with superposition,
but in this case $C \neq 0$
is not required to achieve proper superposition behavior.)

On the other hand, if $S$ is timelike
it may well have a timelike boundary with a non--zero
residue.
More importantly, in the case of
Euclideanized general relativity,
the boundary of {\em any} identification surface
always contributes to the action (up to an overall minus sign)
as though it were timelike and, hence, contributes
to the residue.
In such cases,
can one recover
the superposition law?

The issue whether equation (\ref{super})
strictly requires
invariance of the action
under topological identifications of boundary surfaces
depends on the measure
for the path integral on the left hand side and
its relation to the measure implicit in
the sum over intervening surfaces on the right hand
side.
In the absence of a prescription for either
measure, it is not possible to
say whether invariance of the action
is strictly required.
In fact, the behavior of the action
under spacetime composition
combined with the requirement that
equation (\ref{super})
be obeyed might provide useful insight into
the appropriate measure for quantum gravity.

However, if we adopt the view that invariance of the action
under composition is desirable, then it is appropriate to explore
whether there is
any
special choice for $C$ that renders
the action invariant under spacetime composition.
Let us briefly examine this possibility.

First,
note that if $C$ is truly a functional
of {\em only} the intrinsic geometry of the boundary,
then no choice of $C$ can render the action invariant
under arbitrary spacetime compositions.
If there were such a choice, then $C-\widetilde{C}$
would equal the residue.  Yet, while
$C-\widetilde{C}$ would be a functional of only the intrinsic
geometry of the boundary, the residue
also depends on whether the inner or outer sides of $S$
are identified.  Hence, $C-\widetilde{C}$ could not
equal the residue for arbitrary spacetime compositions.

If one broadens the class of allowed $C$ so that they
can depend not only on the intrinsic geometry of the
boundary but also on its embedding
in some reference spacetime, it
{\em is} possible to ensure invariance of the action
for special classes of spacetime compositions.
One could choose $C$ so that $I$ would actually represent
a `relative action'; that is,
the difference between the `bare action'
and the
action associated with some fixed
reference state satisfying the same
boundary conditions.  [For example, one could
take Minkowski space as the reference state.
This would be equivalent to adopting the Gibbons/Hawking
prescription for $C$~\cite{Gibbons/Hawking}.]
When one varies
the action functional,
the reference state would be kept fixed so
it would have no effect on the equations of motion.
Also, because the
reference action behaves in precisely
the same manner as the bare action under spacetime
composition, invariance of the relative action
would be guaranteed.

Yet, while a given prescription for the reference
state can restore invariance of the action in special cases
of spacetime composition, no
such `relative action' prescription can ever
be defined for arbitrary spacetimes and
arbitrary compositions of them.
The problem, of course, lies in the requirement that
the reference state provide an isometric embedding
of the boundary geometry.  The only state that
can satisfy this requirement for arbitrary
slicings of a given spacetime is the spacetime itself.
Hence, no
{\em independent}
reference state can be defined for arbitrary compositions
of arbitrary spacetimes.

We conclude that while special choices of $C$ may restore
invariance of the action
for special spacetime compositions, it appears that
there is no way to choose $C$ which will restore
invariance for arbitrary spacetime compositions.
In any case, one would hope that if the action
is to be `additive' it should be so regardless
of one's choice of $C$.
This suggests adopting a broader definition of `additivity'.

\section{\label{S5}Generalized Additivity}
Suppose a spacetime
$\{g_{ab},\widetilde{\cal M}\}$ is obtained
from another spacetime
$\{g_{ab}, {\cal M}\}$ either by: (1) topologically identifying
isometric sub--manifolds of ${\cal M}$ to
form a single submanifold ${\cal T}$ of $\widetilde{{\cal M}}$
(see Figure 6),
or (2) by simply removing from ${\cal M}$ a submanifold ${\cal
T}$.
We will say that an action exhibits `generalized additivity'
if
\begin{equation}
I[\widetilde{{\cal M}}]=I[{\cal M}]-I[{\cal T}].
\label{complaw}\end{equation}
We do not require that either ${\cal M}$ or $\widetilde{{\cal
M}}$
be connected and we stress that ${\cal T}$ may be a region of
finite
4--volume. In fact, in the latter case it is easy to show that
equation (\ref{complaw}) holds, essentially because
the contributions from {\em all} the boundaries,
including those of ${\cal T}$,
add up correctly. In other words, whenever identifications
are made on regions of finite 4-volume, generalized
additivity holds.

 From this type of identification we can recover spacetime
composition as defined in Section 1 by going to a limit where
${\cal T}$ collapses on a surface
$S$. The general composition law
then reduces to
\begin{equation}
I[\widetilde{{\cal M}}]=I[{\cal M}]-I[S].
\label{complaw2}\end{equation}
In this case it is possible to show that
\begin{equation}
I[{\cal T}]\to I[S]= {\cal R}[S]+ {\rm
terms~dependent~on}~C.\label{surfact}\end{equation}
In other words, when $C=0$ it is appropriate to interpret
the residue as a {\em finite
action associated with the identification surface} itself.
More generally, when $C\ne0$ we should interpret
the change in the action due to identification of
$S$ as an action associated with this surface.

We conclude that in this generalized sense, the action is always
additive.  The only `unexpected' result is that
when a 4--manifold collapses on a surface with timelike boundary,
its action does not vanish (for arbitrary $C$).

\section{\label{S6}Discussion}

At this point, some comments are in order.
First,
we may not expect the action to be invariant
when topology change is involved.
By identifying one or more boundary surfaces,
we can transform a spacetime of a given topology
to an spacetime of a different topology.
Equation (\ref{resid}) can then be used to determine
how the action of a spacetime changes subject to the
changes in topology.
One particularly important application
of this is in the realm
of gravitational thermodynamics.   In the
Euclidean sector, the
residue associated with placing a boundary
just outside an event horizon
is given by equation (\ref{residerat}).
This residue can be understood as the
entropy associated with the horizon.

Second, we note that
the question of whether or not the action is additive cannot
be divorced from the question of what constraints are to be
imposed along the identification surface and, in particular, at
its
boundary. All the analysis of Sections \ref{S2} and
\ref{S3} supposes that the intrinsic geometry
(and only the intrinsic geometry)
is to be held fixed along $S$.
If different constraints are imposed
along $S$, the behavior of the action under spacetime
composition must be re-examined.

Third, we note that General Relativity is not the only theory
in which the action additivity problem needs to be analyzed and resolved.
Any theory that ascribes a finite action
to the identification surface or its boundary may have
non-vanishing residues. This includes, for example, higher
order gravity theories and non-minimal couplings to other fields.
The results for timelike vs.~spacelike joints, and definition
of generalized additivity, would follow the pattern set in the
Sections above.

As a final note, it is worth observing
that the quasilocal
energy that arises out
of the Hamiltonian formulation of general
relativity~\cite{Brown/York,Hayw/Wong,Hayw/Wong2}
is also non--additive in the `usual sense'
(i.e. it is not invariant under identification
of boundary surfaces).
By exactly repeating the analysis performed
above in one less dimension---that is, by identifying of portions
of the 2--dimensional boundaries of 3--dimensional spaces---one
derives the residues associated with addition of quasilocal
energies.  A more detailed discussion of this phenomenon
will follow
in a separate publication.

{\bf Acknowledgments}\/:
It is a pleasure to thank Bill Unruh, Don Witt,
Kristin Schleich, Jonathan Simon, and Valrey Frolov for valuable
discussions on this subject.  GH is grateful
to the National Science and
Engineering Council of Canada
and the Canadian Institute for Theoretical
Astrophysics for supporting this research.

\eject
\centerline{{\bf List of Figure Captions}}
\begin{enumerate}
\item
(a) Spacetime $\{g^{ab},{\cal M}\}$ with boundary $\{h^{ab},B\}$
is topologically reconfigured to form
a new spacetime $\{g_{ab},\widetilde{{\cal M}}\}$ with boundary $\{h^{ab},
\widetilde{B}\}$.\\
(b) This is achieved by topologically identifying
surfaces $S_1$ and $S_2$ to form a single surface $S$.  The behavior
of the action under the topological identification
is affected by the joints
at $\bar{S}$ (marked in bold) and $\bar{S}^\ast$.
\item
(a) By topologically identifying
$S_1$ and $S_2$, the spacetime $\{g^{ab},{\cal M}\}$
becomes
a new spacetime $\{g_{ab},\widetilde{{\cal M}}\}$.  Note that $\widehat{S}=S$
so $\bar{S}^\ast=0$.\\
(b) at timelike joints $\bar{S}_1$ and $\bar{S}_2$
the unit
normal changes direction by angles $\Theta_1$ and $\Theta_2$, respectively.
After identification, the unit normal changes direction by an angle $\Theta$.
Note that $\Theta=\Theta_1+\Theta_2-\pi$.
\item
(a) The identification surface has topology
$J\times \bar{{\sf I}}$.
A coordinate
$\varepsilon$ parametrizes $\bar{{\sf I}}$
with $-\varepsilon_0<\varepsilon<+\varepsilon_0$.
Thus, $S$ can be decomposed into a
family of 2--surfaces, $J_{(\varepsilon)}$.\\
(b) In the limit that $\varepsilon_0\to 0$,
the identification surface collapses on the 2--surface
$J$.
\item
A diagram algebra is useful
to keep track of the residue for complicated
identification surfaces.  A solid dot represents a joint,
which contributes to the action one quarter
of its area.  An empty bar represents a 3--surface, which contributes
to the action negative one eighth the area of its 2--boundary surfaces.
\item
Topological identification at a spacelike hypersurface with
spacelike boundary.
In the 2-space orthogonal to $\bar{S}_1$,
$n^a_1$ is the outward pointing unit vector tangent to $S_1$,
$u^a_1$ is the future pointing
unit vector tangent to $B - S_1$, and
$\tilde{u}^a_1$ is the future pointing normal to $n^a_1$.
Similarly, $u^a_2$ and $n^a_2$ are, respectively, past and outward pointing
unit normals at $\bar{S}_2$.
\item
A spacetime $\{g_{ab},{\cal M}\}$ is transformed into a new
spacetime $\{g_{ab},\widetilde{\cal M}\}$
by topologically identifying two submanifolds
to form a single region ${\cal T}$.
\end{enumerate}
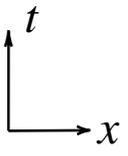
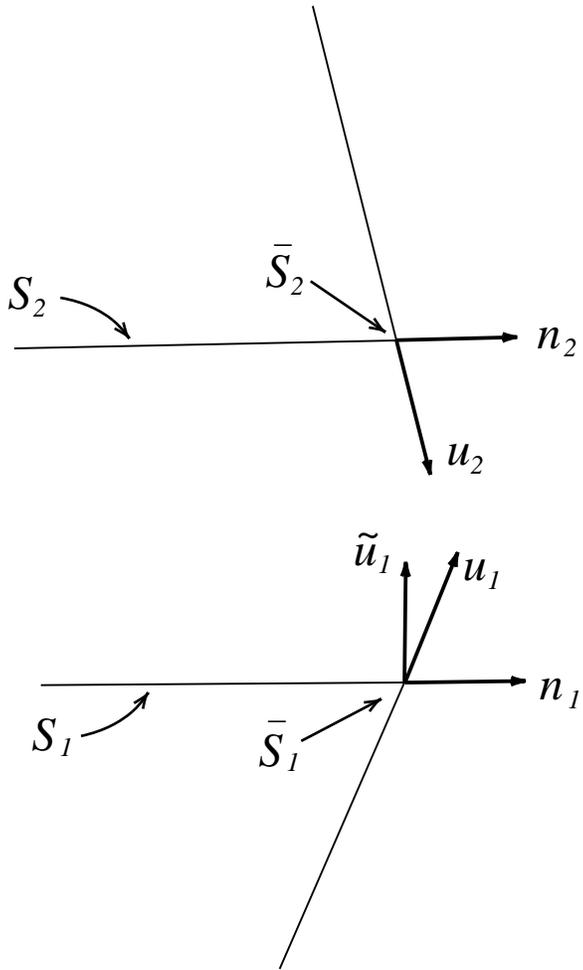
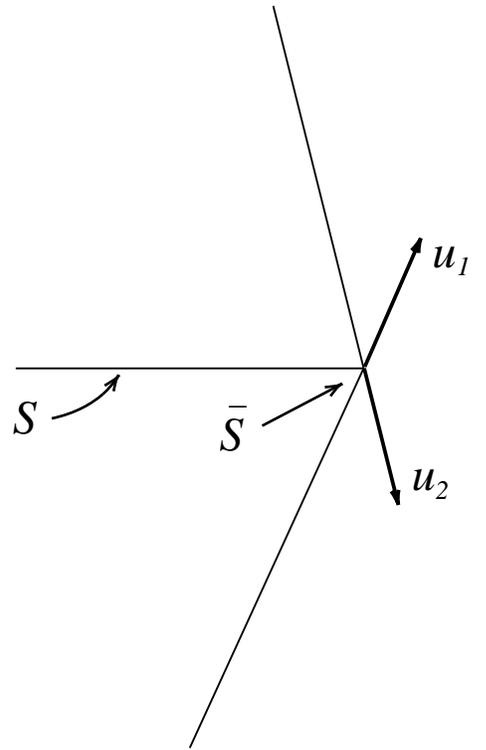
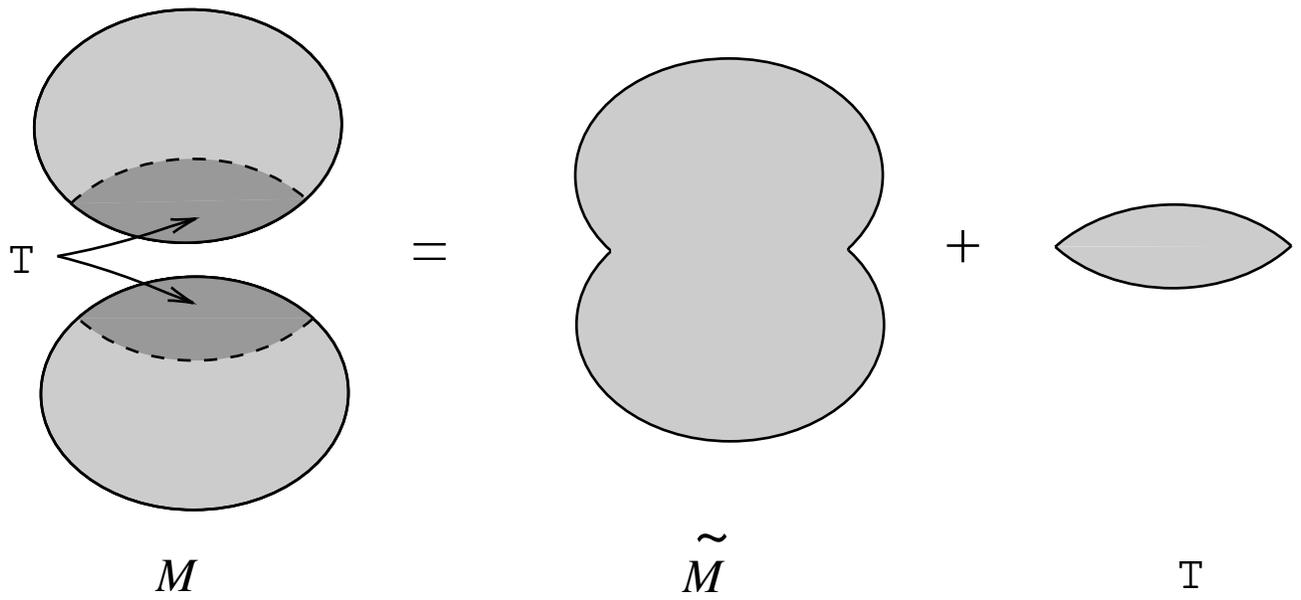
\end{document}